\documentclass[preprint,12pt]{elsarticle}

\usepackage{amssymb}

\usepackage{lineno}
\usepackage[T1]{fontenc}
\usepackage[unicode,colorlinks=true]{hyperref}
\journal{Nucl. Instr. Meth. Phys. Res. A}

\begin{document}

\begin{frontmatter}





\title{ \vspace*{0.11cm} 
Kaonic lead feasibility measurement at DA$\Phi$NE to solve the charged kaon mass discrepancy}



\author[a1]{D. Bosnar \corref{cor1}} \ead{bosnar@phy.hr}
\cortext[cor1]{Corresponding author}
\author[a2]{L. Abbene}
\author[a3]{C. Amsler}
\author[a4]{F. Artibani}
\author[a4]{M. Bazzi}
\author[a5]{M. Bragadireanu}
\author[a2]{A. Buttacavoli}
\author[a3]{M. Cargnelli}
\author[a6]{M. Carminati}
\author[a4]{A. Clozza}
\author[a4]{F. Clozza}
\author[a6]{G. Deda}
\author[a4]{L. De Paolis}
\author[a4,a7]{R. Del Grande}
\author[a4]{K. Dulski}
\author[a7]{L. Fabbietti}
\author[a6]{C. Fiorini}
\author[a1]{I. Fri\v{s}\v{c}i\'{c}}
\author[a4]{C. Guaraldo}
\author[a4]{M. Iliescu}
\author[a8]{M. Iwasaki}
\author[a9]{A. Khreptak}
\author[a1]{M. Makek}
\author[a4]{S. Manti}
\author[a3]{J. Marton}
\author[a9,a10]{P. Moskal}
\author[a4]{F. Napolitano}
\author[a9,a10]{S. Nied\'{z}wiecki}
\author[a11]{H. Ohnishi}
\author[a4,a12] {K. Piscicchia}
\author[a4]{A. Scordo}
\author[a4]{F. Sgaramella}
\author[a4,a5,a12]{D. Sirghi}
\author[a4,a5]{F. Sirghi}
\author[a9,a10]{M. Skurzok}
\author[a9,a10]{M. Silarski}
\author[a4]{A. Spallone}
\author[a11] {K. Toho}
\author[a3]{M. T\"{u}chler}
\author[a4]{O. Vasquez Doce}
\author[a3]{J. Zmeskal}
\author[a4]{C. Curceanu}

\affiliation[a1]{organization={Department of Physics, Faculty of Science, University of Zagreb},
            addressline={Bijeni\v{c}ka c.32}, 
            city={Zagreb},
            postcode={10000}, 
            country={Croatia}}
               
            \affiliation[a2]{organization={Dipartimento di Fisica e Chimica - Emilio Segr\`{e}, Universit\`{a} di Palermo},
            addressline={Viale Delle Scienze, Edificio 18}, 
            city={Palermo},
            postcode={90128}, 
            country={Italy}}

            \affiliation[a3]{organization={Stefan-Meyer-Institut f\"{u}r Subatomare Physik},
            addressline={Kegelgasse 27}, 
            city={Vienna},
            postcode={1030}, 
            country={Austria}}
            
\affiliation[a4]{organization={Laboratori Nazionali di Frascati INFN},
            addressline={Via E. Fermi 54}, 
            city={Frascati},
            postcode={00044}, 
            country={Italy}}
            
              \affiliation[a5]{organization={Horia Hulubei National Institute of Physics and Nuclear Engineering (IFIN-HH)},
            addressline={Reactorului 30}, 
            city={Magurele},
            postcode={077125}, 
            country={Romania}}      
            
   \affiliation[a6]{organization={Politecnico di Milano, Dipartimento di Elettronica, Informazione e Bioingegneria and INFN Sezione di Milano},
            addressline={Via Giuseppe Ponzio 34}, 
            city={Milano},
            postcode={20133}, 
            country={Italy}}

       \affiliation[a7]{organization={Excellence Cluster Universe, Technische Universit\"{a}t M\"{u}nchen Garching},
            addressline={Boltzmann str. 2}, 
            city={Garching},
            postcode={85748}, 
            country={Germany}}
            
         \affiliation[a8]{organization={RIKEN},
            addressline={2-1 Hirosawa, Wako, Saitama}, 
            city={Tokyo},
            postcode={351-0198}, 
            country={Japan}}

    \affiliation[a9]{organization={Faculty of Physics, Astronomy, and Applied Computer Science, Jagiellonian University},
            addressline={\L ojasiewicza 11}, 
            city={Krakow},
            postcode={30-348}, 
            country={Poland}}
            
 \affiliation[a10]{organization={Centre for Theranostics, Jagiellonian University },
            addressline={Kopernika 40}, 
            city={Krakow},
            postcode={31-501}, 
            country={Poland}}     
            
         \affiliation[a11]{organization={Research Center for Electron Photon Science (ELPH), Tohoku University,},
            addressline={1-2-1 Mikamine,Taihaku-ku},
            city={Sendai},
            postcode={982-0826}, 
            country={Japan}}                

   \affiliation[a12]{organization={Centro Ricerche Enrico Fermi - Museo Storico della Fisica e Centro Studi e Ricerche ``Enrico Fermi''},
            addressline={Via Panisperna 89A}, 
            city={Roma},
            postcode={00184}, 
            country={Italy}}

\begin{abstract}

An HPGe detector equipped with a transistor reset preamplifier and readout with a CAEN DT5781 fast pulse digitizer was employed in the measurement of X-rays from kaonic lead at the DA$\Phi$NE $e^+e^-$ collider at the Laboratori Nazionali di Frascati of INFN. A thin scintillator in front of a lead target was used to select kaons impinging on it and to form the trigger for the HPGe detector. We present the results of the kaonic lead feasibility measurement, where we show that the resolution of the HPGe detector in regular beam conditions remains the same as that without the beam and that a satisfactory background reduction can be achieved. This measurement serves as a test bed for future dedicated kaonic X-rays measurements for the more precise determination of the charged kaon mass.

\end{abstract}




\end{frontmatter}


\section{Introduction}

The current value of the charged kaon mass, $m_K=493.677 \pm 0.016$~MeV/$c^2$, was determined as a weighted average of six measurements with very different uncertainties \cite{pdg}. The two most precise measurements \cite{gal, denisov}, which largely determine this value, differ by 60 keV and have uncertainties of approximately 10 keV. This led to a large scaling factor on the error and, consequently, the accuracy on the charged kaon mass (32 p.p.m.) is one order of magnitude worse than the accuracy on the charged pion mass (1.3 p.p.m.). 

The value of the charged kaon mass is important in non-perturbative QCD where its uncertainty has a large influence on the K-N scattering lengths, and consequently on the kaon-nucleon sigma terms, which reflect the degree of chiral symmetry breaking \cite{jaffe}. The accuracy of the charged kaon mass also determines the systematic uncertainty of the $D^0$-meson mass  \cite{lees} which, in turn, has an influence on the determination of the masses of the excited charm mesons \cite{pdg} and mixing parameters in the $D^0$-$\overline{D}^0$ system \cite{lees}. It is therefore imperative to resolve the charged kaon mass discrepancy  and it is sufficient to have the same precision as the previous two most precise measurements 
to substantially improve its accuracy \cite{beer}. 

One of the most precise determinations of the charged kaon mass can be obtained by using the measurements of the X-rays emitted from kaonic atoms \cite{pdg, gal, denisov}. In this approach, one electron in the atom is replaced by a negative kaon and a kaonic hydrogen-like atom is formed. The kaon cascades to lower states of the kaonic atom by emitting X-rays and is eventually absorbed by the nucleus. By measuring the energies of the emitted X-rays and comparing them with the theoretically calculated values for some chosen kaon mass, which is varied until agreement is obtained, the charged kaon mass can be determined. 
It is important to choose X-rays from the cascading spectrum for which the influence of strong interaction between the nucleus and the kaon is totally negligible, and the screening of the inner electrons does not contribute significantly. One also includes important corrections such as the finite nuclear-charge distribution,  vacuum polarization, electron screening and parallel non-circular transitions, as well as higher-order relativistic corrections and static polarization of the orbiting particle and the nucleus, see e.g. \cite{gal}.

In this context we measured the X-rays from kaonic lead with an HPGe detector at the DA$\Phi$NE e$^+$e$^-$ collider at the Laboratori Nazionali di Frascati (INFN-LNF)  to study the feasibility of X-ray measurements from  targets suitable for the determination of the charged kaon mass. 
This was performed in parallel with SIDDHARTA-2 measurements, which investigate X-rays from light kaonic atoms \cite{revmodphys}. 

DA$\Phi$NE provides low momenta kaons and, in contrast to previous measurements which mostly used energetic kaons, a degrader is not needed to slow down the kaons, and there are no secondary hadronic particles in the beam. However, there is background caused by e$^+$ and e$^-$ beam losses and  their interactions with the surroundings, as well as the background originating from the kaons absorbed by the nuclei in the lead target. The latter background type cannot be eliminated, but according to GEANT4 simulations, its contribution can be neglected for our purpose. 

Since there are no detailed simulations of e$^+$ and e$^-$ beam losses, the background they cause can only be determined by measurements in the DA$\Phi$NE hall in real beam conditions. The aim of this study is 
to experimentally assess the impact of the  event rate seen by the HPGe detector on its energy resolution, and to investigate the possibilities to reject the background coming from the  e$^+$ and e$^-$ beams to further reduce the total background.

To process the high rates we used an HPGe detector with a transistor reset preamplifier (TRP) whose signals were fed into one of the four inputs of a CAEN DT5781 fast pulse digitizer unit. To reject the background from the e$^+$ and e$^-$ beams we employed a thin scintillation detector in front of the lead target 
for the detection of kaons impinging on it. 
In addition, the transit times of the kaons and the minimum ionizing particles (MIPs) from the e$^+$e$^-$ interaction point to the scintillation detector were also measured and used for the reduction of background.

The results show that the resolution of the HPGe detector in the beam condition remains the same as that with the beam off. The number of collected events and the observed signal-to-background ratio for X-rays from  kaonic lead also show that a precise determination of the charged kaon mass is feasible in a reasonable time.

In the next section we describe the HPGe setup at DA$\Phi$NE. The third section presents the results of the measurement of the X-rays from  kaonic lead and we conclude with a discussion of the results and an outlook.

\section{The HPGe setup at DA$\Phi$NE}
The HPGe setup was installed together with the SIDDHARTA-2 setup  which measures X-rays from kaonic transitions in gaseous targets with SDD detectors, and whose primary goal is to measure X-rays from kaonic deuterium at DA$\Phi$NE \cite{florin}.

Kaons originate from the decays of $\phi$ mesons produced in e$^+$ and e$^-$  annihilations at the 
interaction point (IP) of the DA$\Phi$NE e$^+$e$^-$ collider \cite{zobov}.
The momenta of kaons are slightly higher toward the inner (boost) than toward the outer (anti-boost) side of the DA$\Phi$NE ring. Their mean momenta are 141 MeV/c and 113 MeV/c, respectively. 

\begin{figure}[ht]
{\includegraphics[width=65mm]{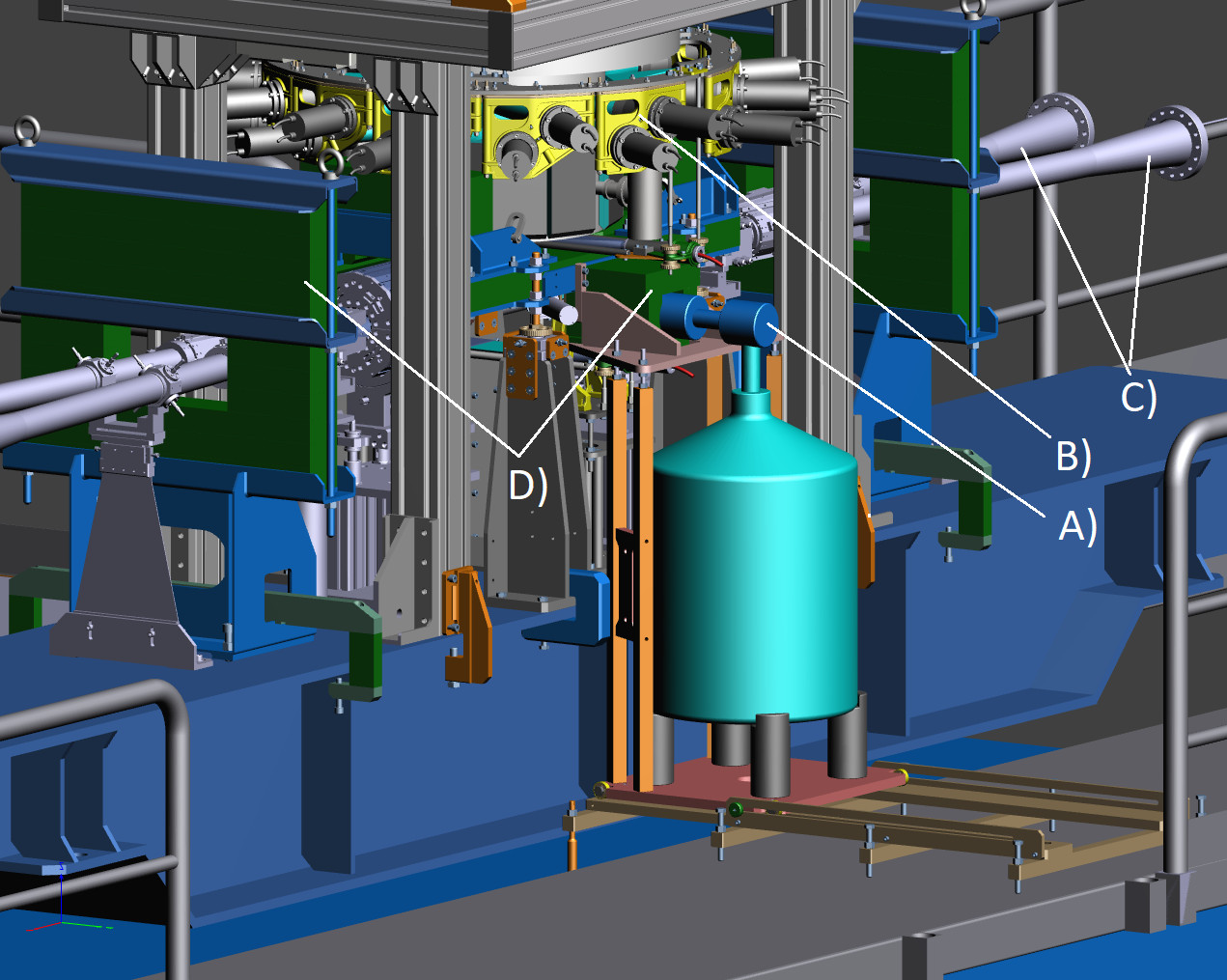}}
\hspace*{1cm}
{\includegraphics[width=65mm]{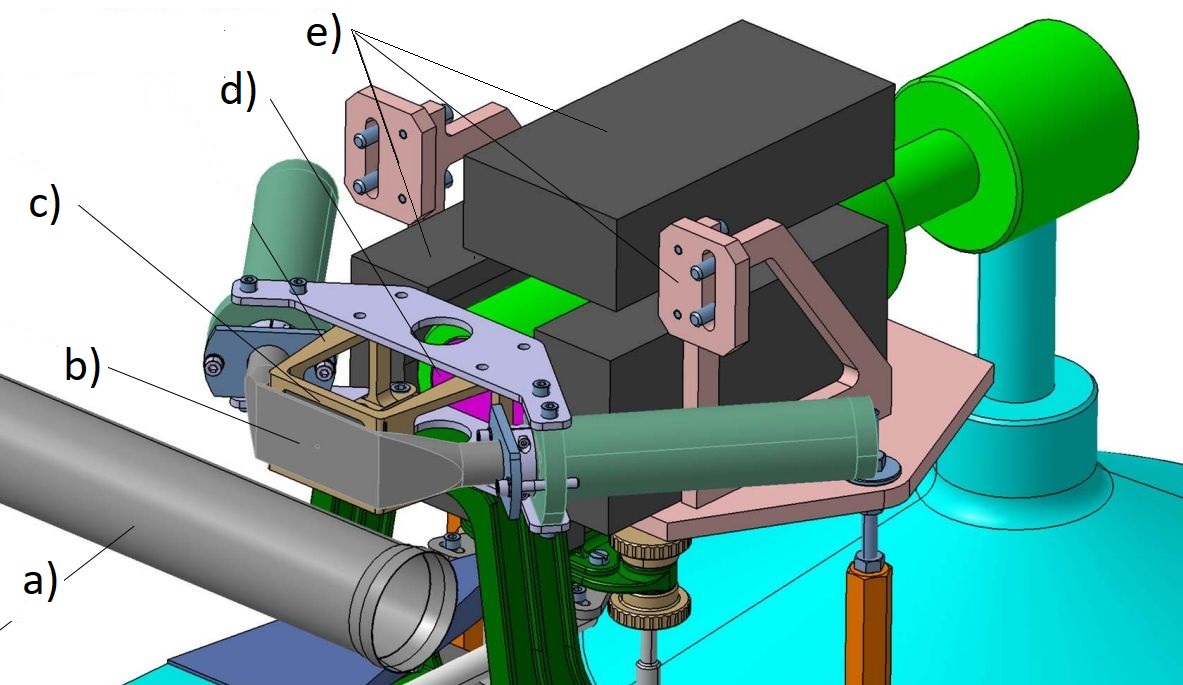}}
\caption{The location of the HPGe detector in the SIDDHARTA-2 setup, left: A) the HPGe detector, B) the SIDDHARTA-2 setup, C) beam pipes, D) shielding. Details of the HPGe setup, right: a) beam pipe, b) scintillation detector SC1, c) lead target, d) active part of the HPGe detector, e) lead shielding with holder.}
\label{hpge-setup}
\end{figure}

The target and SDD detectors of the SIDDHARTA-2 setup are placed above the e$^+$e$^-$ IP, Figure~\ref{hpge-setup}, left. Two plastic scintillators (80~x~40~x~2~mm$^3$ Saint-Gobain Crystals BC-408), SC1 and SC2, which serve as a luminosity monitor in the SIDDHARTA-2 measurements \cite{magda}, are placed on  opposite sides of the IP in the e$^+$e$^-$ beam plane, with the long side parallel to the beams. The signals from each scintillator are read by fast R4998 Hamamatsu PMTs coupled to 6 cm fish-tail plastic light guides on both 40 mm sides of the scintillator.

A lead target of the same size as the scintillation detector SC1 and with a thickness of 1.5 mm  was placed immediately behind.
Its distance from the IP was 78 mm and it completely stopped the entering kaons. They were selected by applying a threshold on the signals from SC1.
To enhance their identification and reduce the background, the trigger signal for the HPGe detector was formed as a coincidence between SC1, SC2 and the DA$\Phi$NE radio-frequency (RF) signal  ($\simeq$368 MHz) divided by 4, using a CAEN N455 coincidence unit module. The usage of the RF/4 signal is a consequence of the hardware limitation, given that the constant fraction discriminator units (Ortec 935 CFD) are restricted to working with signals below 200 MHz.



The HPGe detector is a p-type detector produced by Baltic Scientific Instruments. The active part is 
a cylinder with a base diameter of 59.8 mm and a height of 59.3 mm. The front side of the cylinder could be positioned at a minimal distance of 155 mm from the lead target, a constraint posed by the geometry of the SIDDHARTA-2 setup, and the measurement was performed at this position. The active part of the HPGe detector was shielded by 5 cm thick lead bricks, Figure~\ref{hpge-setup}, right. The front brick (not shown in the figure) was 2.5 cm thick with a circular hole of 4 cm in diameter in the center, for the X-rays to reach the active part of the detector from the lead target. 

\begin{figure}[h]
{\includegraphics[width=130mm]{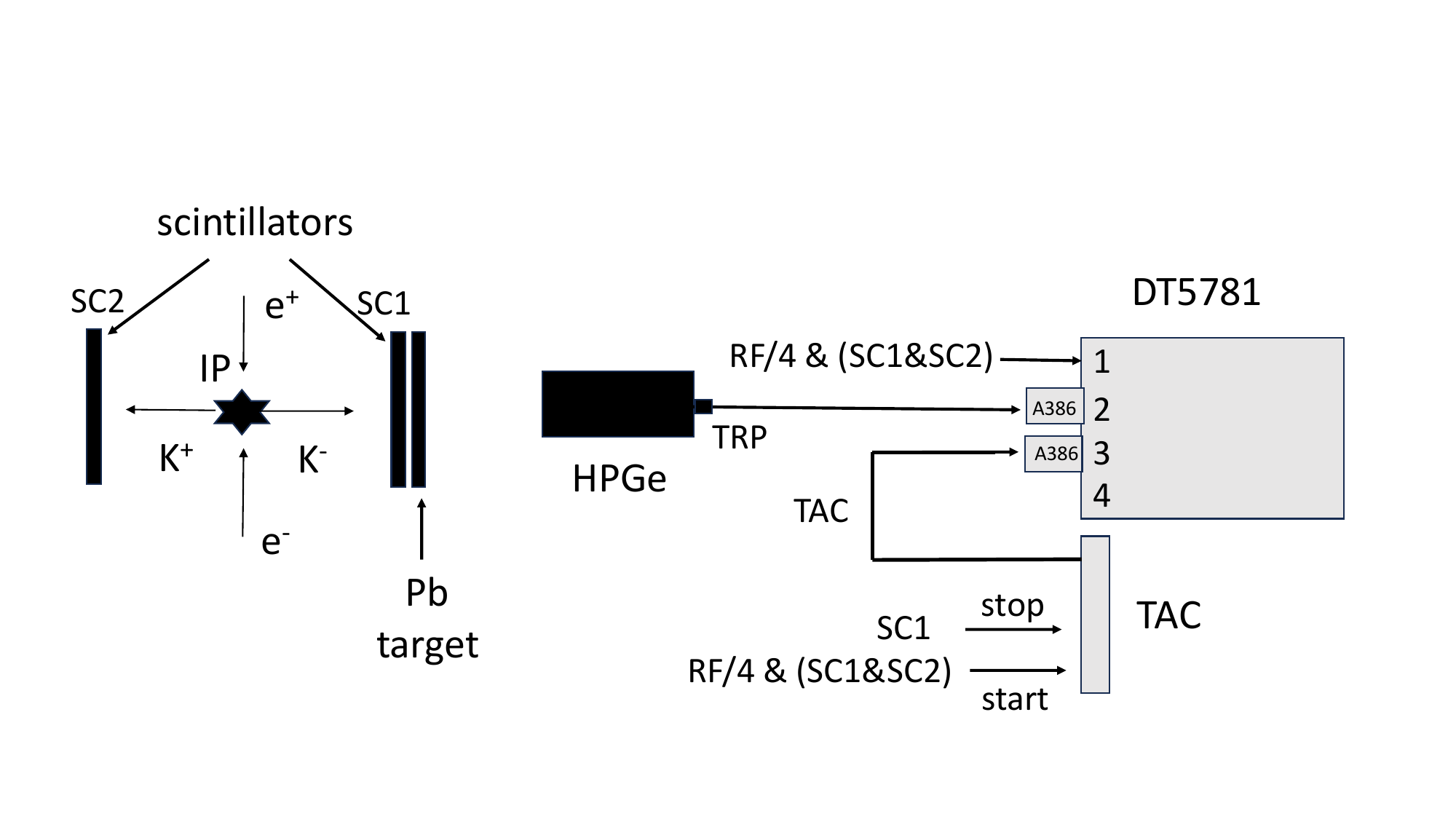}}
\caption{Schematics of the HPGe detector system (not to scale, dimensions are given in the text) and electronics for the data acquisition.}
\label{shema}
\end{figure}

A schematic representation of the HPGe detector system and electronics for the data acquisition is shown in Figure~\ref{shema}.

Signals from the TRP were fed into a CAEN DT5781 fast pulse digitizer through the CAEN AC coupler A386. 
The DT5781 unit has 4 signal inputs with 100 MS/s 14-bit waveform digitizers and is capable of selecting coincidence events  which come in the selected inputs in the predetermined coincidence time window. 
 In the measurement of kaonic lead X-rays, signals from the TRP were put in  coincidence with the trigger signals, which opened a broad coincidence gate of 2.5 $\mu$s.

Since it is not possible to entirely discriminate kaons from MIPs with a threshold on the scintillator signal height, their transit times from the IP to the scintillation detector SC1 were measured as well. 
An ORTEC 566 time-to-amplitude-converter (TAC) was used for this purpose. The start signal for the TAC was given by the trigger signal.
As the stop signal, a signal produced by a CAEN N93B dual mean timer unit was used, into which  discriminated signals (by ORTEC 935 CFDs) from both sides of the scintillation detector SC1 were fed.
The output signals from the TAC were put into the DT5781 unit (by using the CAEN A386 AC coupler) and were taken in coincidences with the corresponding events in the HPGe.
These transit times were used in the off-line analysis to additionally discriminate between the kaons and the MIPs.

The CAEN MC$^2$ Analyzer software  was used for  data acquisition, on-line analysis and control. In particular, it was used for the reconstruction of
signal amplitudes in the HPGe detector with the trapezoid algorithm \cite{trapezoid} tuned for high rates, and for the determination of the signal occurrence time, which can be done with a resolution of 10 ns. 
These data were stored in binary format for off-line analysis, which was performed using the CERN Root software \cite{root}.

\section{Measurement of kaonic lead X-rays with the HPGe detector at DA$\Phi$NE}
The measurement 
was performed during the SIDDHARTA-2 experiment in June 2023. The total integrated luminosity measured with the SIDDHARTA-2 luminosity monitor was 39.4 pb$^{-1}$.

The calibration of the energy scale of the digitizer was done by using a 1~$\mu$Ci $^{133}$Ba source. This was also used to determine and monitor the energy resolution of the detector during the whole period of the measurement.
\begin{figure}[h]
\includegraphics[width=130mm]{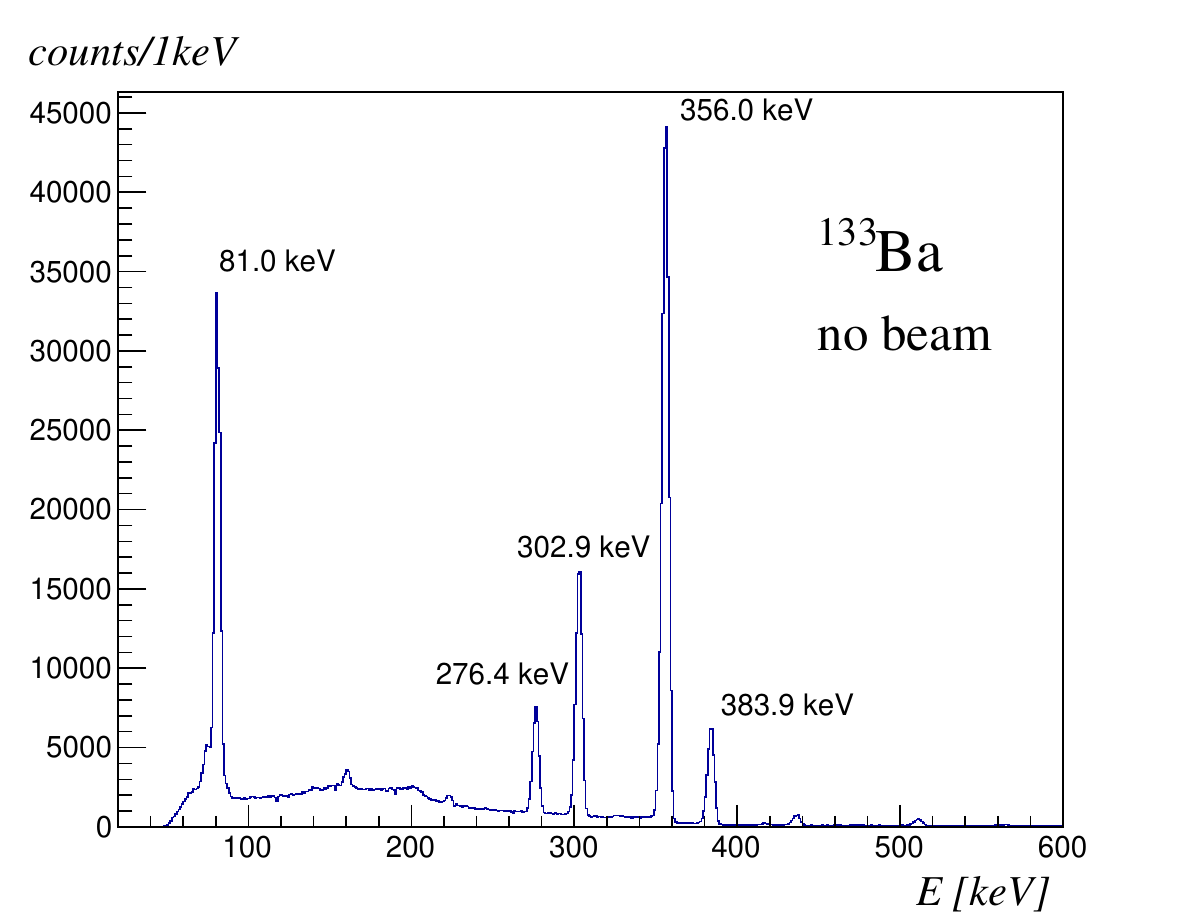}
\makebox[0pt][r]{
    \raisebox{11em}{%
      \includegraphics[width=.23\linewidth]{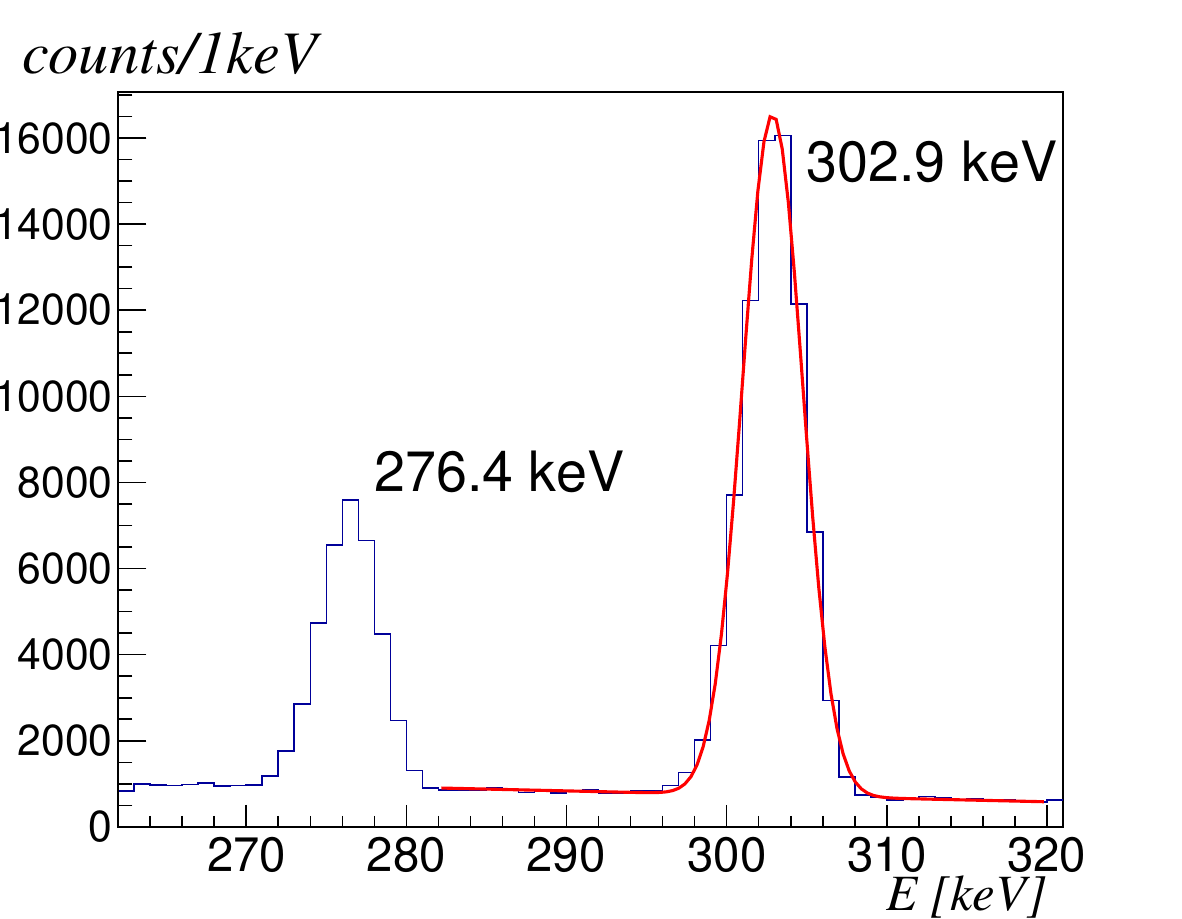}
    }\hspace*{15em} }%
\caption{The spectrum of $\gamma$-rays from the $^{133}$Ba source with beams off. The inset shows the peak at 302.9 keV fitted with a Gaussian and a linear function for the background estimation. The energy resolution is 4.39 $\pm$ 0.02 keV (FWHM).}
\label{133Ba-302-1}
\end{figure}

\begin{table}[h]
 \centering
 \caption{The $^{133}$Ba lines in terms of peak, positions and energy resolutions  obtained from a fit of the energy-calibrated spectrum, and residuals (R).}
 \vspace{0.5cm}
 \begin{tabular}{c| c | c | c }
 $^{133}$Ba line  & Peak position  & Resolution (FWHM)  & R (10$^{-4}$) \\
  (keV) & (keV) & (keV) &  \\
 \hline
  80.9971 & 80.994 $\pm$ 0.006 & 3.74$\pm$ 0.01 & 0.38 \\
 276.398 & 276.41 $\pm$ 0.01 & 4.30 $\pm$ 0.02 & -0.36 \\
  302.853 & 302.868 $\pm$ 0.008 & 4.39 $\pm$ 0.02 & -0.49\\
  356.017 & 355.893 $\pm$ 0.005 & 4.582 $\pm$ 0.007 & 3.48 \\
  383.851 & 383.76 $\pm$ 0.01 & 4.70 $\pm$ 0.03 & 2.42 \\
 \end{tabular}
\end{table}

A spectrum of $^{133}$Ba taken in the DA$\Phi$NE hall with beams off at the beginning of the measurement and in the same detector configuration as for kaonic lead X-rays  is shown in Figure~\ref{133Ba-302-1}. The energy resolution at 302.9 keV was 4.39 $\pm$ 0.02 keV (FWHM), the inset of Figure~\ref{133Ba-302-1}, and it remained stable during the measurement. This line is of special interest since it is the closest to the ($9\rightarrow8$) transition line in kaonic lead. The positions and energy resolutions of the other peaks from $^{133}$Ba obtained from a fit of the energy-calibrated spectrum, as well as their relative deviations with respect to the nominal values (residuals, R), are shown in Table 1. 

\begin{figure}[h]
{\includegraphics[width=70mm]{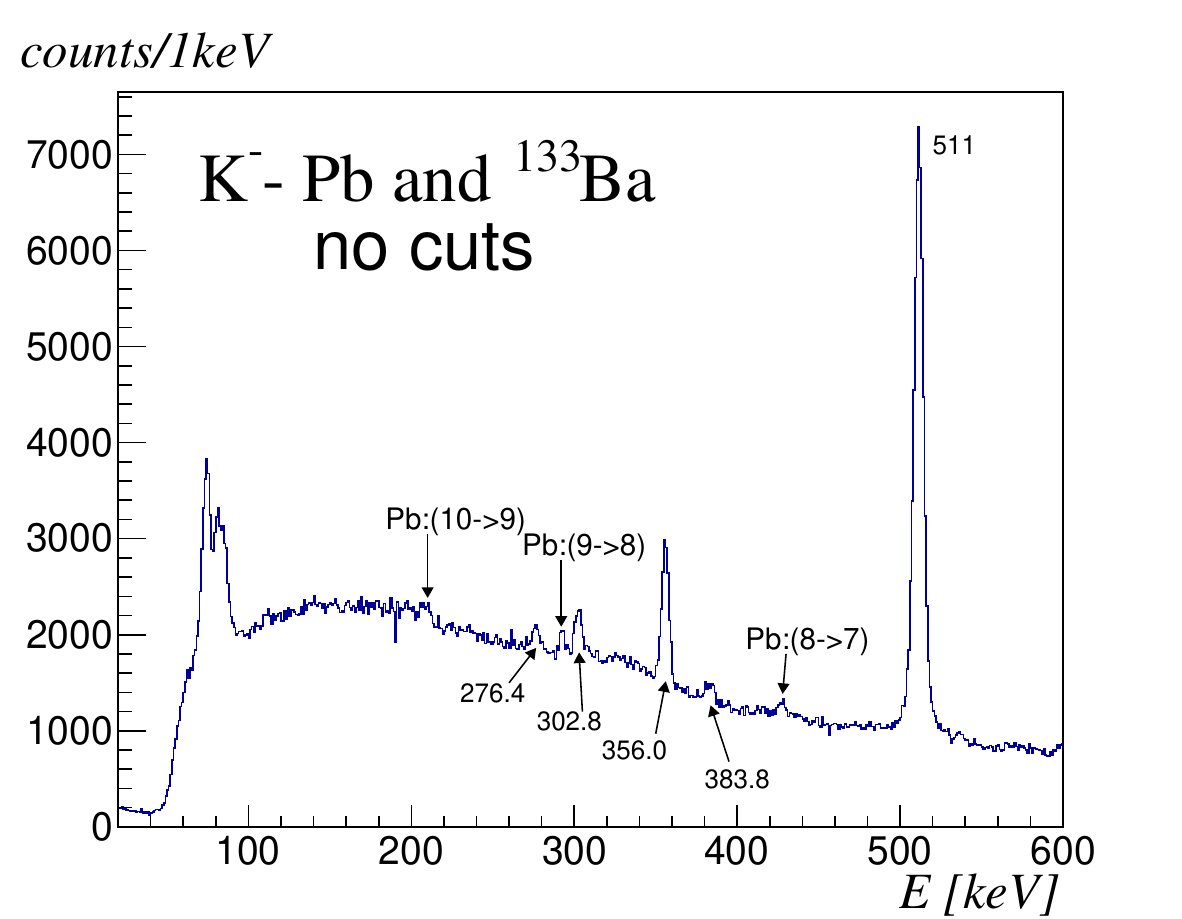}}
{\includegraphics[width=70mm]{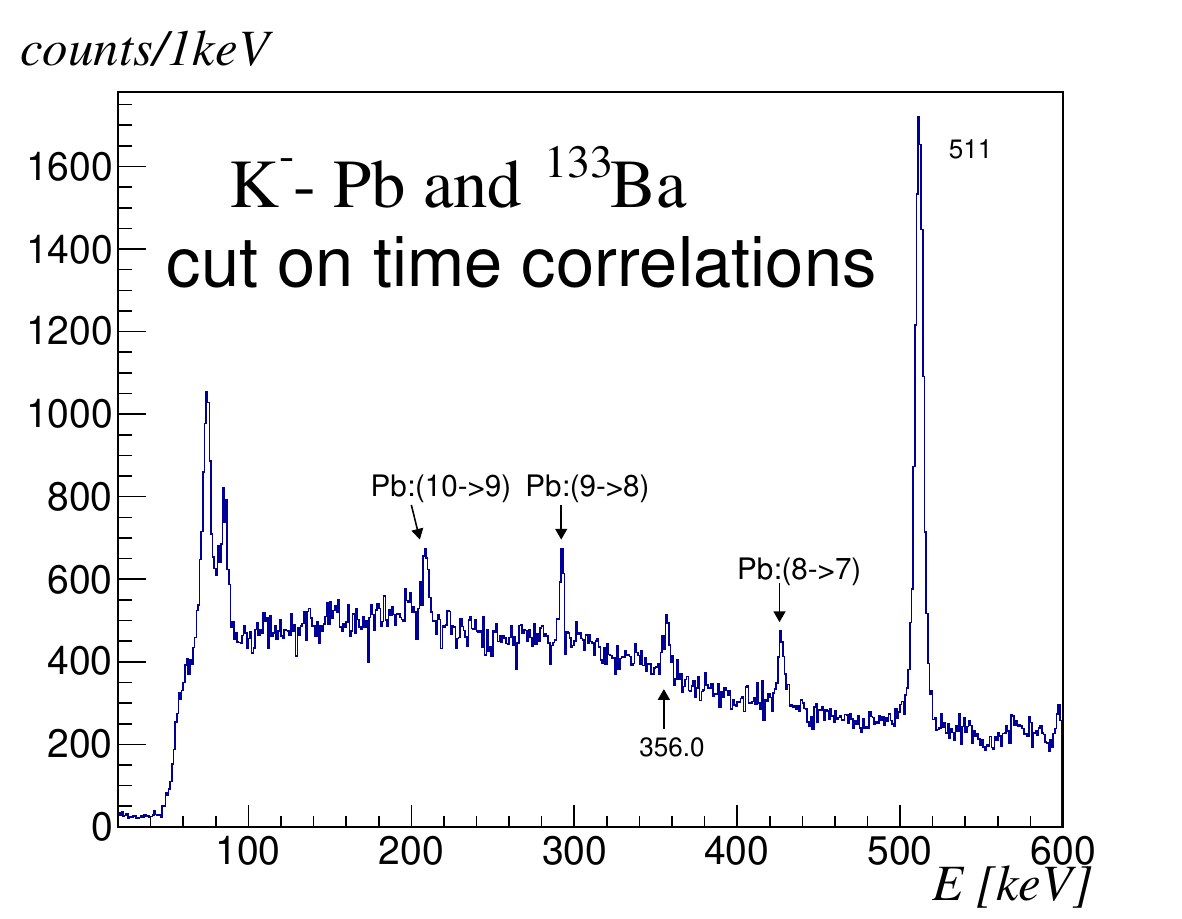}}
\caption{The energy spectrum seen by the HPGe detector with a 2.5~$\mu$s coincidence gate, left, and the spectrum with a cut of 400 ns on the time difference between signals in the scintillation detector SC1 and HPGe, right. }
\label{hpge-all}
\end{figure}

In the measurement of kaonic lead X-rays, signals from the HPGe were taken in coincidence with signals from the scintillation detector SC1 and with the corresponding signals from the TAC by using the DT5781 unit.
The spectrum of events seen by the HPGe detector in coincidence with the scintillation detector SC1 within a broad coincidence gate of 2.5 $\mu s$ is shown in Figure~\ref{hpge-all}, left.

\begin{figure}[h]
\begin{center}
{\includegraphics[width=70mm]{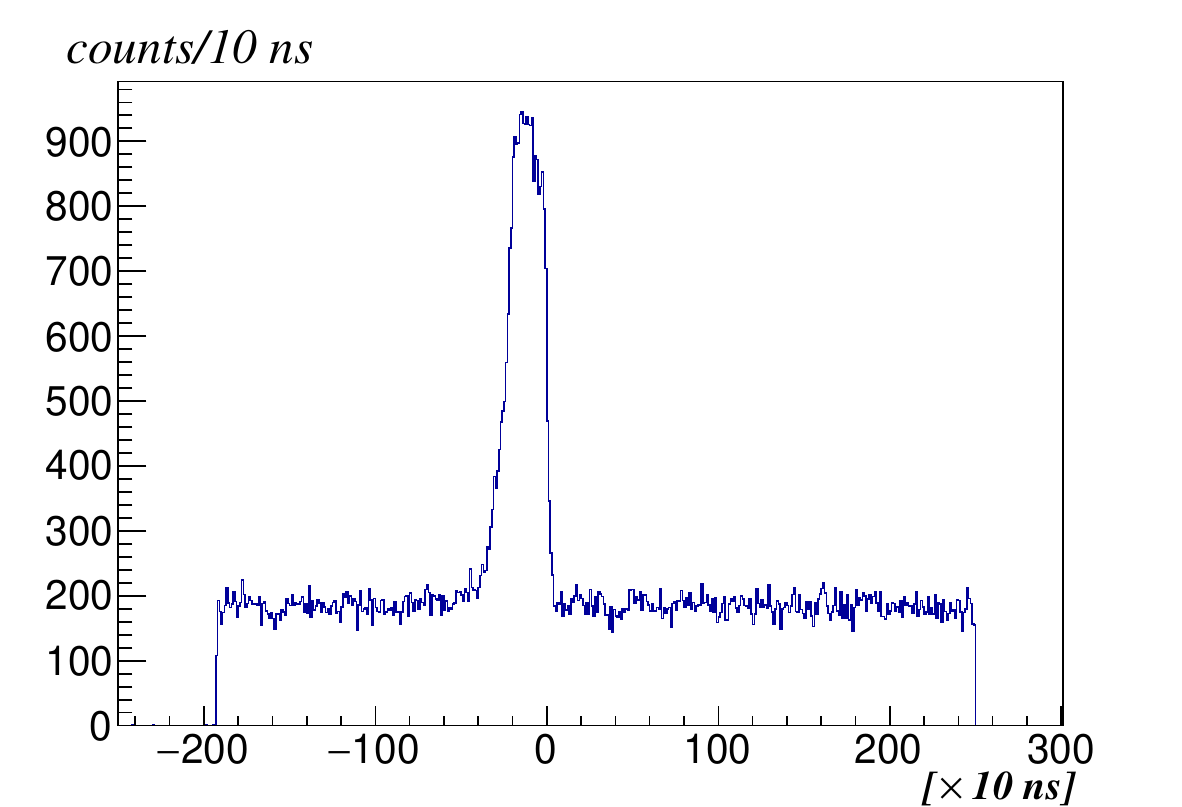}}
\end{center}
\caption{Time correlation between the signals in the scintillation detector SC1 and HPGe, determined by the digitizer (time resolution of 10 ns). }
\label{t3t1}
\end{figure}

In the off-line analysis, a narrower cut of 400 ns on the time difference between signals in the scintillator and HPGe determined by the digitizer was imposed. An example of this time correlation is shown in Figure~\ref{t3t1} while the spectrum obtained with this cut is shown in Figure~\ref{hpge-all}, right.

\begin{figure}[h]
{\includegraphics[width=120mm]{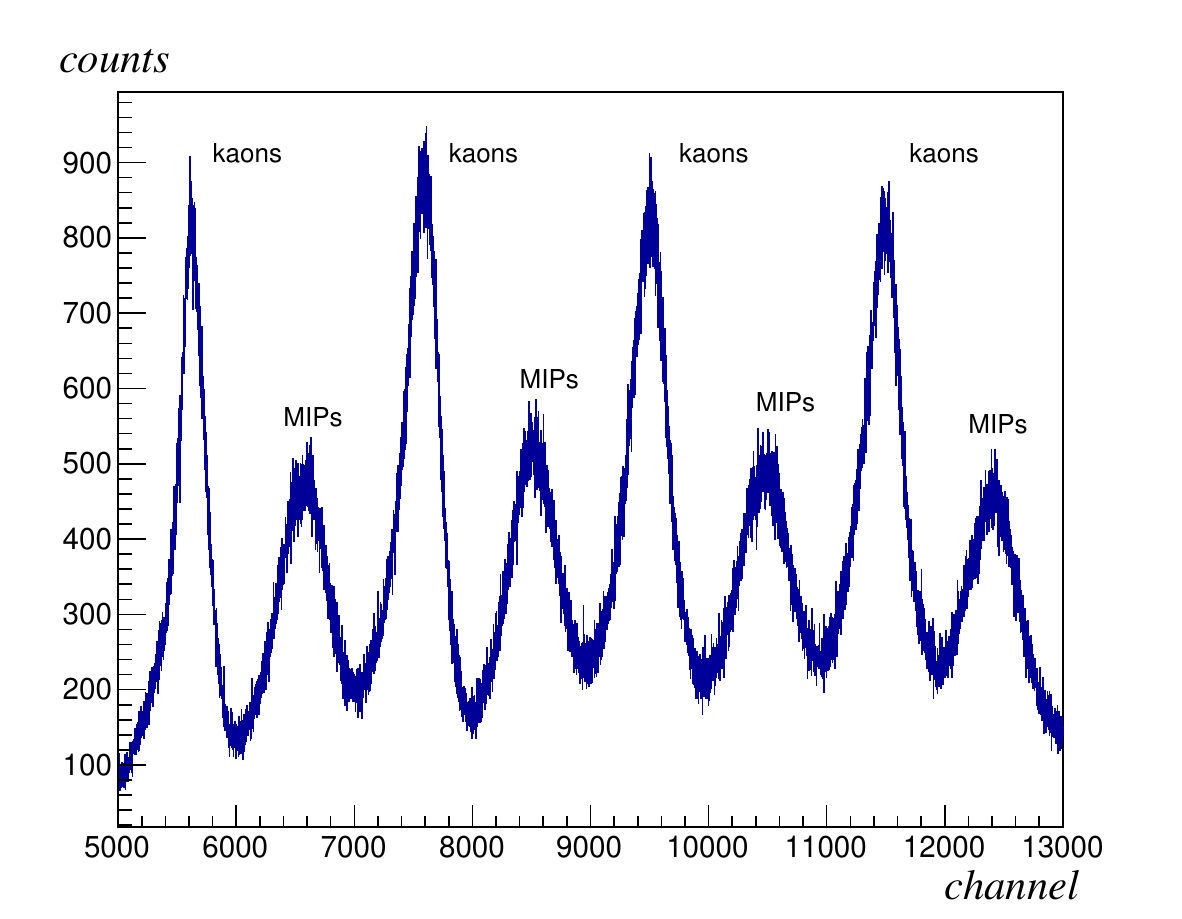}}
\caption{Spectrum of the transit times (expressed in channels of the digitizer) of the  kaons and MIPs from the IP to the scintillation detector SC1.}
\label{TAC}
\end{figure}

A further reduction of the background was obtained by using the transit times of the kaons and MIPs from the IP to the scintillation detector SC1,  measured by the TAC and expressed in channels of the digitizer (Figure~\ref{TAC}). The transit time difference between  the MIPs and kaons is approximately 2.7 ns.
The 8 peak structures, 4 peaks from the kaons and 4 from the MIPs, are a consequence of
the usage of the RF/4 signal \cite{magda, ale}. 

\begin{figure}[h]
\includegraphics[width=130mm]{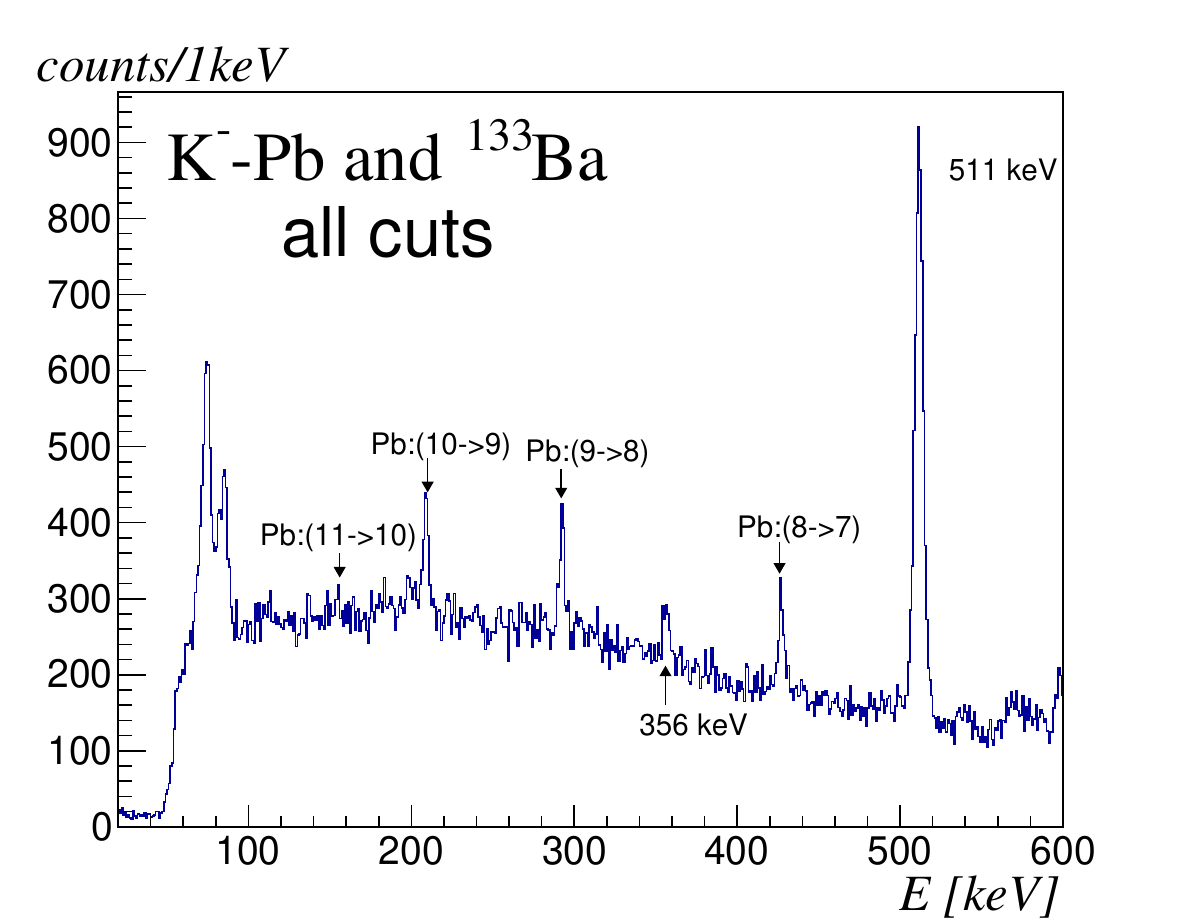}
\makebox[0pt][r]{
    \raisebox{13em}{%
      \includegraphics[width=.25\linewidth]{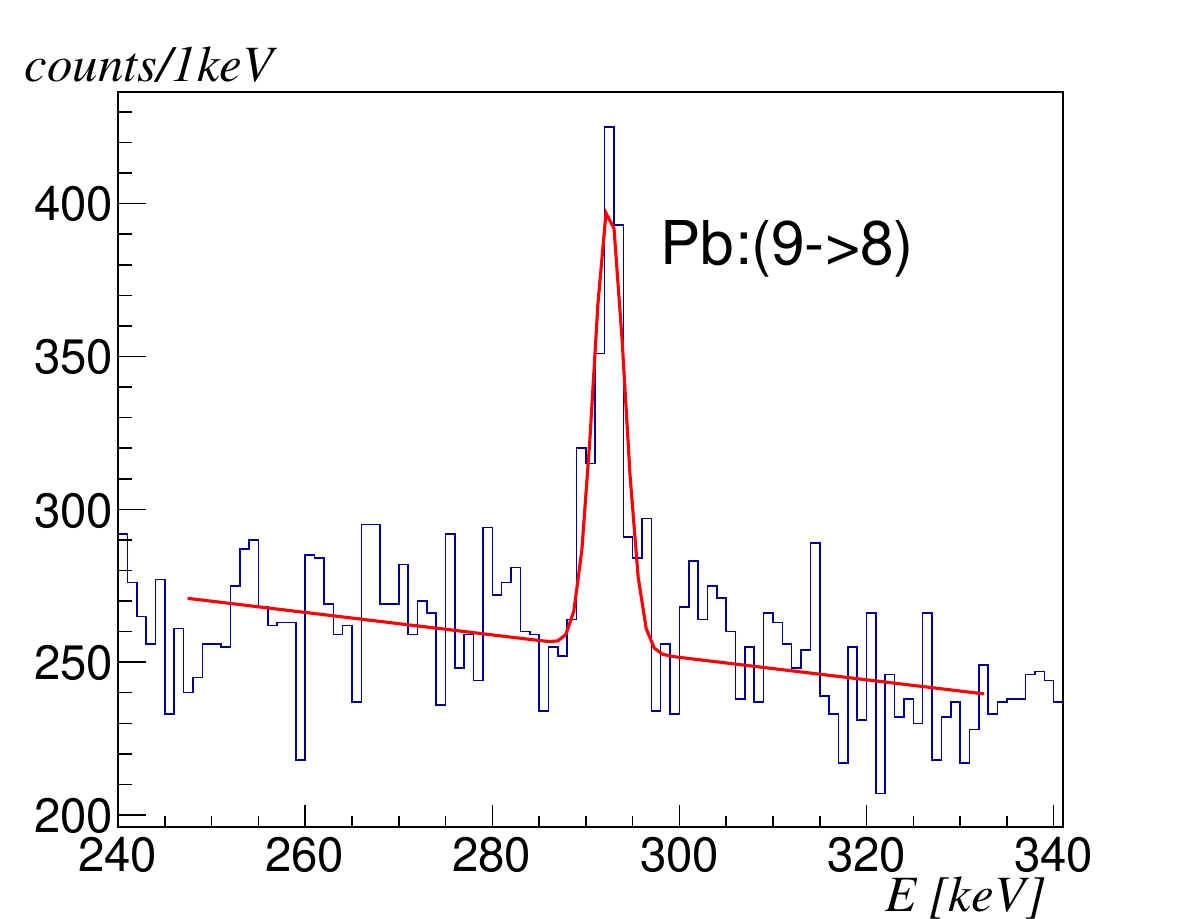}
    }\hspace*{8.0em} }%
\caption{The spectrum seen by the HPGe detector after applying all cuts. The inset shows the peak at 292.47 $\pm$ 0.17 keV with a fit done by a Gaussian and a linear function for the background, the energy resolution is 3.97 $\pm$ 0.49 keV (FWHM).}
\label{hpge-all-cuts}
\end{figure}

Figure~\ref{hpge-all-cuts} shows the final energy spectrum seen by the HPGe detector after applying all the cuts. 
 Clear peaks are visible at 208.92 $\pm$ 0.17 keV, 292.47 $\pm$ 0.17 keV and 427.07 $\pm$ 0.24 keV which come from the (10$\rightarrow$9), (9$\rightarrow$8) and 
 (8$\rightarrow$7) transitions in the kaonic lead, respectively. The peak at 154.2 $\pm$ 1.2 keV, from the (11$\rightarrow$10) transition, is less pronounced.
 
 Besides the peak at 511 keV which comes from positron annihilation,
 there are peaks which come from the transitions in ordinary lead with which the sensitive part of the HPGe detector was shielded. Expected peaks are at 72.80 keV and 74.96 keV, which are not resolved due to the resolution of the detector, and the peak at 84.94 keV.
 
 There are also visible peaks at 356.0 keV and 81.0 keV   which come from the $^{133}$Ba source, which was  in front of the sensitive part of the HPGe detector during the measurement. The other peaks from  $^{133}$Ba are less pronounced in this spectrum and are more visible in the spectrum shown in Figure ~\ref{hpge-all}, left, where no cuts were applied.

\section{Discussion and outlook}

The inset of Figure~\ref{hpge-all-cuts} shows the peak at 292.47 $\pm$ 0.17 keV from (9$\rightarrow$8) transitions in kaonic lead. The energy resolution is 3.97 $\pm$ 0.49 keV. Compared  with the resolution  of 4.39 $\pm$ 0.02 keV at 302.9 keV of $^{133}$Ba in the measurement with  the beams off, Table 1, it can be concluded that there is no worsening of the resolution in measurements in the full beam conditions at DA$\Phi$NE. 

 This feasibility measurement was performed with the HPGe detector whose energy resolution at the 302.9 keV $^{133}$Ba line was 4.39 keV  in the configuration with the DT5781 fast pulse digitizer. 
 This is worse than the standard resolution (approximately 1.3 keV, FWHM) of such  detectors at this energy and in the same configuration. 
 
Assuming the standard resolution of the HPGe detector, we made an estimation of the required number of events in the (9$\rightarrow$8) transition peak  to reach the 10~keV accuracy on the charged kaon mass.
In a simple approach, we used the Rydberg formula for kaonic lead atoms, with the kaon-nucleus reduced mass, to obtain the relation between the precision of the transition 
and the accuracy of the charged kaon mass. To reach the 10~keV accuracy, the required precision of the (9$\rightarrow 8$) transition needs to be approximately 6 eV.
This estimation is also confirmed by the preliminary results of a more sophisticated calculation based on the self-consistent multiconfiguration Dirac-Fock method \cite{desc, indelicato}, which is in progress. 
Approximately 8500 events in the  (9~$\rightarrow$~8) transition peak are needed to reach the 10 keV accuracy on the charged kaon mass, assuming a detector with a resolution of 1.3 keV at 292 keV.


%
%
The total number of events in the  (9~$\rightarrow$~8) peak in our measurement is 770 $\pm$ 65. This implies that a total integrated luminosity of 435 pb$^{-1}$ is needed to achieve the required accuracy by using only the (9 $\rightarrow$ 8) transition in kaonic lead.

\begin{table}
 \centering
 \caption{Summary of the clearly visible X-rays from kaonic lead.}
 \vspace{0.5cm}
 \begin{tabular}{c| c | c | c }
 K$^-$-Pb transition & Peak position  & Resolution (FWHM) & Number of events \\ 
 & (keV) & (keV) &  \\ \hline
 10 $\rightarrow$ 9 & 208.92 $\pm$ 0.17 & 3.68 $\pm$ 0.42 & 584 $\pm$ 30 \\
  9 $\rightarrow$ 8 & 292.47 $\pm$ 0.17 & 3.97 $\pm$ 0.49 & 770 $\pm$ 65 \\
  8 $\rightarrow$ 7 & 427.07 $\pm$ 0.24 & 4.37 $\pm$ 0.54 & 457 $\pm$ 45 \\
 \end{tabular}
\end{table}
However, there are clearly visible peaks at 208.92 keV and 427.07 keV with  584 $\pm$ 30  and 457 $\pm$ 45 events in the peak, respectively (Table 2), which can also be used to determine the charged kaon mass more precisely.



\section{Acknowledgements}
We thank C. Capoccia from LNF-INFN for his 
contribution in designing and building the holder for the HPGe detector and the holder for the lead target. We also thank the DA$\Phi$NE staff for the excellent working  conditions provided and for their support. 

Part of this work was supported by the Croatian Science Foundation under project IP-2022-10-3878; the EU STRONG-2020 project (grant agreement No. (824093); the Austrian Science Fund (FWF): [P24756-N20 and P33037-N]; the EU Horizon 2020 project under the MSCA G.A. 754496; Ministero degli Affari Esteri e dela Cooperazione Internazionle, EXOTICA project, PO21MO03; the Polish Ministry of Science and Higher Education grant No. 7150/E-338/M/2018; the Polish National Agency for Academic Exchange (grant No. PPN/BIT/2021/1/00037); the SciMat and qLife Priority Research Areas budget under the program Excellence Initiative - Research University at the Jagiellonian University.
This publication was made possible through the support
of Grant 62099 from the John Templeton Foundation. The opinions
expressed in this publication are those of the authors and do not
necessarily reflect the views of the John Templeton Foundation.



\clearpage

\begin{thebibliography}{00}

\bibitem{pdg} R.L. Workman, et al. (Particle Data Group), Review of Particle Physics, Prog. Theor. Exp. Phys. 2022 (2022) 083C01, https://doi.org/10.1093/ptep/ptac097.  
\bibitem{gal} K.P. Gall, et al., Precision Measurements of the $K^-$ and $\Sigma^-$ Masses, Phys. Rev. Lett. 60 (3) (1988) 186-189, doi.org/10.1103/PhysRevLett.60.186.
\bibitem{denisov} A.S. Denisov, et al., New measurements of the mass of the K$^-$ meson, JETP Lett. 54 (10) (1991) 558-563.
\bibitem{jaffe} R.L. Jaffe, et al., The pattern of chiral symmetry breaking and the strange quark content of the proton, Comments Nucl. Part. Phys. 17 (1987) 163-175.
\bibitem{lees} J.P. Lees, et al., Measurement of the mass of the $D^{0}$ meson, Phys. Rev. D 88 (2013) 071104, https://doi.org/10.1103/PhysRevD.88.071104.
\bibitem{beer} G. Beer, et al., A new method to obtain a precise value of the mass of the charged kaon, Phys. Lett. B535 (2002) 52-58, https://doi.org/10.1016/S0370-2693(02)01664-7.
\bibitem{revmodphys} C. Curceanu, et al., The modern era of light kaonic atom experiments, Rev. Mod. Phys. 91 (2019) 025006-1, doi:10.1103/RevModPhys.91.025006.
\bibitem{florin} F. Sirghi, et al., Kaonic atoms with SIDDHARTA-2 at the DA$\Phi$NE collider, EPJ Web Conf. 291 (2024) 01008, https://doi.org/10.1051/epjconf/202429101008.
\bibitem{zobov} M. Zobov, et al., Test of 
crab-waist collisions at DA$\Phi$NE $\Phi$ factory, Phys. Rev. Lett. 104 (2010) 174801, doi:10.1103/PhysRevLett.104.174801.
\bibitem{magda} M. Skurzok, et al., Charcterization of the SIDDHARTA-2 luminosity monitor, JINST 15 (10) (2020) P10010, doi:10.1088/1748-0221/15/10/P10010.
\bibitem{trapezoid} V.T. Jordanov, et al., Digital techniques for real-time pulse shaping in radiation measurements, Nucl. Instr. Meth. Phys. Res. A353 (1994) 261-264, https://doi.org/10.1016/0168-9002(94)91652-7.
\bibitem{root} R. Brun and F. Rademakers, ROOT - An object oriented data analysis framework, Nucl. Instr. Meth. Phys. Res. A389 (1997) 81-86, https://doi.org/10.1016/S0168-9002(97)00048-X.
\bibitem{ale} A. Scordo, et al., CdZnTe detectors tested at the DA$\Phi$NE collider for future kaonic atoms measurements, Nucl. Instr. Meth. Phys. Res. A1060 (2024) 169060, https://doi.org/10.1016/j.nima.2023.169060.
\bibitem{desc} J.P. Desclaux, A multiconfiguration relativistic Dirac-Fock program, Comput. Phys. Communication 9 (1975) 31-45, https://doi.org/10.1016/0010-4655(75)90054-5. 
\bibitem{indelicato} O. Gorceix, et al., Multiconfiguration Dirac-Fock studies of two-electorn ions: I. Electron-electron interaction, J. Phys. B: Atom. Mol. Phys. 20 (1987) 639-649, DOI 10.1088/0022-3700/20/4/006.
\end{thebibliography}


\end{document}